\documentclass[]{article}

\usepackage[margin=1in]{geometry}
\usepackage[colorlinks=true,linkcolor=blue,urlcolor=blue,citecolor=blue]{hyperref}

\usepackage{authblk}

\usepackage{amssymb}
\usepackage{amsfonts}
\usepackage{amsmath}
\usepackage{graphicx}
\usepackage{booktabs}
\usepackage{makecell}
\usepackage{xcolor}
\usepackage{caption}
\usepackage{subcaption}
\usepackage{titlesec}

\begin{document}

\title{Non-exotic traversable wormholes in Einstein-Chern-Simons gravity}

\author[1,2]{Mauricio Cataldo\thanks{
		\href{mailto:mcataldo@ubiobio.cl}
		{\texttt{mcataldo@ubiobio.cl}}}}

\author[ ]{Fernando Gom\'{e}z\thanks{
		\href{mailto:fpgomezdiaz@gmail.com}
		{\texttt{fpgomezdiaz@gmail.com}}}}
	
\author[3]{Paola Meza\thanks{
		\href{mailto:paola.meza@uss.cl}
		{\texttt{paola.meza@uss.cl}}}}
	
\author[3]{Cristian Quinzacara\thanks{
		\href{mailto:cristian.quinzacara@uss.cl}
		{\texttt{cristian.quinzacara@uss.cl}}}}	
	
\author[4,5]{Patricio Salgado\thanks{
		\href{mailto:patsalgado@unap.cl}
		{\texttt{patsalgado@unap.cl}}}}	

\affil[1]{Departamento de Física, Universidad del Bío-Bío, Casilla 5-C, Concepción, Chile}
\affil[2]{Centro de Ciencias Exactas, Universidad del Bío-Bío, Casilla 447, Chillán, Chile}

\affil[3]{Departamento de Ciencias Exactas, Facultad de Ingeniería, Universidad San Sebastián, Concepción, Chile}

\affil[4]{Facultad de Ciencias, Universidad Arturo Prat, Avda. Arturo Prat 2120, Iquique 1110939, Chile}
\affil[5]{Instituto de Ciencias Exactas y Naturales (ICEN), Universidad Arturo Prat, Avda. Playa Brava 3256, Iquique 1110939, Chile}

\date{}

\maketitle

\begin{abstract}
We investigate traversable wormhole solutions in five-dimensional Einstein-Chern-Simons (EChS) gravity, a gauge-theoretic extension of General Relativity that introduces a higher-curvature correction parametrized by the combination $\alpha l^2$, where $\alpha$ is a dimensionless coupling and $l$ is a length scale. Working with the Morris-Thorne metric and an anisotropic fluid, we derive exact expressions for the energy density, radial pressure, and lateral pressure at the wormhole throat. We show that the radial Null Energy Condition (NEC) is satisfied at the throat if and only if $\alpha l^2\leq-r_0^2$, being saturated at $\alpha l^2=-r_0^2$ and strictly satisfied for $\alpha l^2<-r_0^2$, independently of the shape function. In the latter regime, the energy density is strictly positive and the full NEC, Weak, Strong, and Dominant Energy Conditions can be simultaneously satisfied. These results are established analytically for arbitrary shape functions and illustrated concretely for the power-law family $b(r)=r_0(r_0/r)^n$, $\Phi=0$, for which all four standard energy conditions are satisfied at the throat for all $n>0$ when the EChS coupling is sufficiently negative. We further show that the radial pressure is negative and geometrically fixed at the throat, independently of $\alpha$, $b(r)$, and $\Phi(r)$. Using the Volume Integral Quantifier (VIQ), we establish that the EChS correction reduces the magnitude of the negative integrated NEC contribution relative to GR, and derive an exact closed-form critical coupling $\alpha l^2_{\rm crit}$ at which the VIQ vanishes. The hierarchy $\alpha l^2_{\rm crit}<-r_0^2<0$ shows that NEC satisfaction at the throat and a non-negative global VIQ are compatible but distinct conditions, both achievable within EChS gravity. All results reduce continuously to those of GR in the limit $l\to0$, confirming the consistency of the framework.
\end{abstract}

\section{Introduction}

The seminal works of Morris and Thorne~\cite{Morris:1988cz} and Morris, Thorne, and Yurtsever~\cite{morris1} established the central role of energy-condition violation in traversable wormhole physics. In General Relativity (GR), the flare-out condition requires matter that violates the Null Energy Condition (NEC), i.e., exotic matter with $\rho+p_r<0$ at the throat. This result follows directly from the flare-out condition imposed on the wormhole geometry, which forces the combination $\rho + p_r$ to be negative when the gravitational dynamics are governed solely by the Einstein field equations~\cite{Visser:1995cc}. Since no known classical matter source violates the NEC, the need for exotic matter has long been regarded as one of the most serious obstacles to the physical viability of wormhole solutions.

Motivated by this fundamental difficulty, a significant body of research has been devoted to exploring whether the requirement of exotic matter can be avoided or substantially reduced within the framework of modified theories of gravity. The central idea is that, in such theories, the gravitational field equations acquire additional geometric terms that contribute to the effective stress-energy tensor supporting the wormhole geometry. As a result, the burden of NEC violation can be transferred from the matter sector to the geometric sector, allowing the physical matter content to satisfy the energy conditions.

Among the earliest and most studied frameworks in this direction is $f(R)$ gravity, where the Einstein-Hilbert action is generalized by replacing the Ricci scalar $R$ with an arbitrary function $f(R)$. Traversable wormhole solutions in $f(R)$ gravity have been widely investigated, and it has been shown that for suitable choices of the function $f(R)$ and the shape function, the matter threading the wormhole can satisfy the NEC and WEC, with the higher-order curvature derivative terms being responsible for sustaining the wormhole geometry~\cite{Lobo:2009ip, Capozziello:2012hr, Radhakrishnan:2024sym}.

A natural and well-motivated extension of GR to higher curvature theories is provided by Gauss--Bonnet and Lovelock gravity. In the context of Einstein--Gauss--Bonnet (EGB) gravity, it has been shown that both static and dynamical wormhole solutions can be obtained, with the higher-order curvature corrections modifying the energy-condition requirements imposed on the matter sector~\cite{Maeda:2008nz,Mehdizadeh:2015jra,Mehdizadeh:2020egj}. In third-order Lovelock gravity, wormhole solutions can be constructed for which the matter threading the wormhole satisfies the energy conditions, with the higher-order curvature terms playing a central role in sustaining the geometry~\cite{Zangeneh:2015jda}. More recently, traversable wormholes were constructed in dilatonic Einstein-scalar-Gauss-Bonnet (EsGB) theory in four spacetime dimensions without requiring any form of exotic matter, demonstrating that EsGB theories always feature such solutions due to the interplay between the scalar field and the quadratic Gauss-Bonnet term~\cite{Antoniou:2019awm}. Similarly, non-exotic wormhole solutions in 4D EGB gravity were found to satisfy the NEC throughout the entire spacetime for the negative branch of solutions~\cite{Mishra:2022pra}.

In symmetric teleparallel gravity, or $f(Q)$ theory, where the gravitational interaction is described by the non-metricity scalar $Q$, traversable wormhole solutions satisfying all four standard energy conditions have been obtained using a power-law model $f(Q) = a(-Q)^n$~\cite{Lu:2024fQ}. A related study demonstrated that a wide variety of wormhole solutions for which the matter fields satisfy energy conditions are accessible in the $f(Q)$ framework, with the presence of $f(Q)$ gravity alone being sufficient to sustain the wormhole without exotic matter~\cite{Rastgoo:2024fQ}. A comparative analysis between $F(Q,T)$ and $F(R,T)$ gravity further revealed that, for identical shape functions, the energy density is consistently positive in both theories, while the radial pressure can be positive in the $F(Q,T)$ framework, in contrast to the always-negative radial pressure found in $F(R,T)$ gravity~\cite{Mustafa:2024comparative}.

It is important to note, however, that a recurring subtlety in this literature concerns the distinction between the matter sector and the effective stress-energy tensor. In most modified gravity frameworks, what satisfies the energy conditions is the effective combination of matter and geometric contributions, not the bare matter alone. As noted in~\cite{Antoniou:2019awm}, even in EsGB theories the NEC is violated in some region near the throat for single-throat wormholes; the violation arises not from the matter but from the synergy between the scalar field and the Gauss-Bonnet term. Furthermore, as emphasized in the review of~\cite{Radhakrishnan:2024sym}, the radial pressure at the wormhole throat remains negative in the majority of modified gravity frameworks due to the geometric origin of the flare-out condition, which is independent of the theory of gravity.

The preceding landscape of modified gravity theories shares a common motivation: the search for a gravitational framework capable of sustaining traversable wormholes without resorting to classically forbidden matter. From a complementary perspective, the unification of fundamental interactions suggests that the gravitational field itself should admit a gauge-theoretic description. The Standard Model of particle physics is built upon gauge theories whose fundamental field is a connection, whereas General Relativity is formulated in terms of a metric. This structural asymmetry motivates the construction of gauge theories for gravity whose fundamental field is likewise a connection. One prominent candidate is the Chern-Simons gravity action proposed by Chamseddine~\cite{cham1,cham2,cham3}, which is a well-defined gauge theory for the gravitational field. However, the presence of higher powers of the curvature in the Chern-Simons action makes its dynamics significantly different from that of standard Einstein-Hilbert gravity.

A key development in reconciling these two pictures was achieved in Refs.~\cite{salg1,edel,concha}, where it was shown that standard five-dimensional General Relativity can be recovered from the so-called Einstein-Chern-Simons (EChS) gravity in an appropriate limit of the coupling constant $l \to 0$, while keeping the effective Newton constant fixed. The EChS framework therefore provides a gauge-theoretic extension of GR that reduces smoothly to the Einstein-Hilbert dynamics and introduces a higher-curvature correction parametrized by a dimensionless coupling $\alpha$. This correction, which has the structure of a Gauss-Bonnet term, modifies the gravitational field equations in a way that is absent in standard GR and opens the possibility of qualitatively new wormhole solutions. The EChS theory has been applied to cosmology~\cite{gomez2,Crisostomo1,gomez1}, black holes~\cite{Quinzacara1}, and stellar equilibrium~\cite{Quinzacara2}, but its implications for wormhole physics have not yet been explored.

In the present work, we investigate traversable wormhole solutions within Einstein--Chern--Simons gravity, exploiting the higher-curvature correction controlled by the parameters $\alpha$ and $l$ introduced by the EChS framework. We show that the radial NEC at the throat, $\rho+p_r\geq0$, is satisfied for $\alpha l^2\leq-r_0^2$, being saturated at $\alpha l^2=-r_0^2$ and strictly satisfied for $\alpha l^2<-r_0^2$, independently of the shape function. In the latter regime, the energy density is strictly positive and the full NEC, WEC, SEC, and DEC can be simultaneously satisfied for suitable choices of the metric functions, even though the radial pressure $p_r$ is always negative. This negativity is a geometric result derived from the throat field equations and is independent of the EChS coupling and the metric functions at the throat. This provides a physically transparent setting in which the standard energy conditions are evaluated directly for the ordinary energy-momentum tensor associated with the vielbein, rather than for an effective stress-energy tensor combining matter and higher-curvature contributions.

The remainder of this paper is organized as follows. Section~\ref{sec:MT} introduces the five-dimensional Morris--Thorne geometry in Einstein--Chern--Simons gravity and derives the matter variables for the sector considered here. Section~\ref{sec:EC} analyzes the standard pointwise energy conditions at the wormhole throat and identifies the role of the EChS coupling in their satisfaction. Section~\ref{sec:powerlaw} specializes the discussion to a power-law family of shape functions, including graphical analyses, the dependence of the throat conditions on $\alpha l^2$, the isotropic-pressure case, and the Volume Integral Quantifier. Finally, Section~\ref{sec:conclusions} summarizes the main results and outlines possible extensions. The complete EChS field equations and their Einstein-like reformulation are collected in Appendices~\ref{app:EChS-field-equations} and \ref{app:EChS-Einstein-like}.

\section{Morris--Thorne wormholes in Einstein--Chern--Simons gravity}\label{sec:MT}

\subsection{Einstein--Chern--Simons gravity}
\label{sec:EChS-gravity}

Einstein--Chern--Simons gravity is a five-dimensional gauge theory obtained through the S-expansion of the AdS algebra \cite{salg1}. In addition to the vielbein $e^{a}$ and the spin connection $\omega^{ab}$, the theory includes the bosonic one-form fields $h^{a}$ and $k^{ab}$. Its gravitational action can be written as 
\begin{equation}
	S_{\mathrm{EChS}}[e,\omega,h,k]
	=
	\frac{1}{8\kappa}
	\int_{M}
	\epsilon_{abcde}
	\left(
	\alpha l^{2}R^{ab}R^{cd}e^{e}
	+
	\frac{2}{3}R^{ab}e^{c}e^{d}e^{e}
	+
	2l^{2}k^{ab}R^{cd}T^{e}
	+
	l^{2}R^{ab}R^{cd}h^{e}
	\right),
	\label{eq:EChS-action-main}
\end{equation}
where wedge products between differential forms are understood. Here, $R^{ab}$ and $T^{a}$ are the curvature and torsion two-forms, respectively, $\kappa$ is the five-dimensional gravitational coupling constant, $l$ is a length parameter, and $\alpha$ is dimensionless. In the limit $l\rightarrow0$, with $\kappa$ fixed, the non-Einstein contributions vanish and the Einstein--Hilbert action is recovered.

We restrict to the torsionless sector, $T^{a}=0$, set $k^{ab}=0$, and assume the matter source carries no intrinsic spin. Under these conditions, the vielbein field equation becomes
\begin{equation}
	\frac{1}{8}
	\epsilon_{abcde}
	\left(
	2R^{bc}e^{d}e^{e}
	+
	\alpha l^{2}R^{bc}R^{de}
	\right)
	=
	-\kappa \hat T_{ba}\star e^{b},
	\label{eq:EChS-wormhole-field-equation}
\end{equation}
where $\hat T_{ab}$ denotes the energy-momentum tensor of the physical matter, and $\star$ is the Hodge dual operator\footnote{For an oriented orthonormal coframe $\{e^{a}\}$ in a five-dimensional spacetime, the action of the Hodge operator on a basis $p$-form is 
\begin{equation*} 
	\star \left ( e^{a_{1}}\protect \cdots e^{a_{p}} \right ) = \protect \frac {1}{(5-p)!} \epsilon ^{a_{1}\protect\cdots a_{p}}{}_{a_{p+1}\protect \cdots a_{5}} e^{a_{p+1}}\protect \cdots e^{a_{5}}, 
\end{equation*} 
where wedge products are understood and the	indices of the Levi--Civita symbol are raised with the $\eta$-metric.\label{HodgeDual}}.

The second term on the left-hand side of Eq.~\eqref{eq:EChS-wormhole-field-equation} represents the EChS higher-curvature correction. Equation~\eqref{eq:EChS-wormhole-field-equation} is the field equation relevant for the wormhole analysis developed in the main text. The remaining field equations of the complete EChS system, together with their consistency conditions and Einstein-like reformulation, are presented and discussed in Appendices~\ref{app:EChS-field-equations} and \ref{app:EChS-Einstein-like}.

\subsection{Field equations for the Morris--Thorne geometry}
\label{sec:Morris-Thorne-field-equations}

We consider the five-dimensional static, spherically symmetric Morris--Thorne line element
\begin{equation}
	\mathrm{d}s^{2}=-e^{2\Phi(r)}\mathrm{d}t^{2}
	+
	\frac{\mathrm{d}r^{2}}{1-b(r)/r}
	+
	r^{2}\mathrm{d}\Omega_{3}^{2},
	\label{eq:Morris-Thorne-metric}
\end{equation}
where $\Phi(r)$ is the redshift function, $b(r)$ is the shape function, and $\mathrm{d}\Omega_{3}^{2}$ is the line element of the unit three-sphere. The wormhole throat is located at $r=r_{0}$ and satisfies the condition $b(r_{0})=r_{0}$, while the shape function must also satisfy the flare-out condition $b'(r_{0})<1$.

In an orthonormal frame, the physical matter is described by the anisotropic energy-momentum tensor 
\begin{equation}
	\hat T^{a}{}_{b}
	=
	\operatorname{diag}
	\left(
	-\rho,
	p_{r},
	p_l,
	p_l,
	p_l
	\right),
	\label{eq:matter-tensor}
\end{equation}
where $\rho(r)$ is the energy density, $p_{r}(r)$ is the radial pressure, and $p_l(r)$ is the lateral pressure.

Substituting Eqs.~\eqref{eq:Morris-Thorne-metric} and \eqref{eq:matter-tensor} into
Eq.~\eqref{eq:EChS-wormhole-field-equation} gives 
\begin{subequations}
	\label{eq:matter-components}
	\begin{align}
		\kappa\rho(r)&=\frac{3\big(b(r)+rb'(r)\big)}{2r^{3}}
		+\frac{3\alpha l^{2}b(r)\big(-b(r)+rb'(r)\big)}{2r^{6}},
		\label{eq:energy-density}
		\\
		\kappa p_{r}(r)&=-\frac{3\left(r\Phi'(r)b(r)-r^{2}\Phi'(r) +b(r)\right)}{r^{3}}
		-\frac{3\alpha l^{2}\Phi'(r)b(r)\big(b(r)-r\big)}{r^{5}},
		\label{eq:radial-pressure}
		\\
		\kappa p_l(r)&=\frac{r-b(r)}{r}	\left(\Phi''(r)+\Phi'(r)^{2}\right)
		+\frac{\big(4r-3b(r)-rb'(r)\big)\Phi'(r)}{2r^{2}}
		-\frac{b'(r)}{r^{2}}\notag
		\\
		&\quad+\frac{\alpha l^{2}}{r^{5}}\left\{b(r)r\big(r-b(r)\big)	\left(\Phi''(r)+\Phi'(r)^{2}\right)+\Phi'(r) \left(r-\frac{3}{2}b(r)\right)
		\big(-b(r)+rb'(r)\big)\right\}.
		\label{eq:lateral-pressure}
	\end{align}
\end{subequations}

Each equation contains a five-dimensional Einstein contribution and an EChS correction proportional to $\alpha l^{2}$. The latter describes the modification produced by the EChS curvature-squared term.

In the limit $l\rightarrow0$, these corrections vanish and Eqs.~\eqref{eq:matter-components} reduce to the five-dimensional Einstein equations for the Morris--Thorne geometry. They coincide with Eqs.~(15)--(17) of Ref.~\cite{Cataldo1} for $N=4$.

\section{Energy conditions}\label{sec:EC}

Evaluating Eqs.~\eqref{eq:matter-components} at the throat $r=r_{0}$ gives
\begin{subequations}\label{eq:emtem}
	\begin{align}
		\kappa \rho\big|_{r_0} &= \frac{3}{2r_0^2}\left(1+b'(r_0)\right)
		+\frac{3\alpha l^2}{2r_0^4}\left(-1+b'(r_0)\right), \label{ec:rho_throat} \\
		\kappa p_{r}\big|_{r_0} &= -\frac{3}{r_{0}^{2}}, \label{ec:pr_throat} \\
		\kappa p_{l}\big|_{r_0} &=\frac{\left(1-b'(r_{0})\right)
			\Phi'(r_{0})}{2 r_{0}}\left(1+\frac{\alpha l^{2}}{r_{0}^{2}}\right)
		-\frac{b'(r_{0})}{r_{0}^{2}}, \label{ec:pl_throat}
	\end{align}
\end{subequations}
where primes denote derivatives with respect to $r$, evaluated at the throat $r = r_0$. From Eq.~(\ref{ec:pr_throat}), the radial pressure at the throat is always negative, regardless of the values of $\alpha$, $r_0$, or the shape function.

We now analyze in detail each of the standard pointwise energy conditions at the throat.

\subsection{Null Energy Condition (NEC)}

The NEC requires $\rho + p_r \geq 0$ and $\rho + p_l \geq 0$ for all null vectors. Using Eqs.~\eqref{eq:emtem}, the radial combination, denoted by NEC$_1$, is
\begin{equation}
	\kappa(\rho + p_{r})\Big|_{r_0}=\frac{3}
	{2r_{0}^{2}}\big(-1+b'(r_{0})\big)\left(1+\frac{\alpha l^{2}}{r_{0}^{2}}\right). \label{ec:NEC1}
\end{equation}
The flare-out condition requires $b'(r_0)<1$, and therefore $b'(r_0)-1<0$. Consequently, the sign of $\rho+p_r$ is determined by the factor $1+\alpha l^2/r_0^2$. Three cases arise:
\begin{itemize}
	\item If $\alpha l^2>-r_0^2$, then $\rho+p_r<0$, and NEC$_1$ is violated.
	\item If $\alpha l^2=-r_0^2$, then $\rho+p_r=0$, and NEC$_1$ is saturated.
	\item If $\alpha l^2<-r_0^2$, then $\rho+p_r>0$, and NEC$_1$ is satisfied.
\end{itemize}
In particular, the Einstein limit $\alpha=0$ belongs to the first case, so the radial NEC$_1$ is necessarily violated, as expected in five-dimensional Einstein gravity.

For the lateral combination, denoted by NEC$_2$, one obtains
\begin{equation}
	\kappa(\rho+p_l)\Big|_{r_0}=\frac{1}{2r_0^2}\left\{4+\big(1-b'(r_0)\big)
	\left[r_0\Phi'(r_0)\left(1+\frac{\alpha l^2}{r_0^2}\right)-	1-\frac{3\alpha l^2}{r_0^2}\right]\right\}.
	\label{ec:NEC2}
\end{equation}
In the regime $\alpha l^2<-r_0^2$, the flare-out condition implies
$1-b'(r_0)>0$, while $(1+\alpha l^2/r_0^2)<0$ and $(-1-3\alpha l^2/r_0^2)>0$. Thus, the EChS contribution independent of the redshift derivative gives a positive contribution to NEC$_2$. The contribution proportional to $\Phi'(r_0)$, however, has the opposite sign to $\Phi'(r_0)$. Consequently, $\Phi'(r_0)<0$ strengthens NEC$_2$, whereas $\Phi'(r_0)>0$ weakens it.

In particular, if $\Phi'(r_0)\leq0$, NEC$_2$ is automatically satisfied. If $\Phi'(r_0)>0$, NEC$_2$ imposes an upper bound on the redshift derivative. Therefore, there exists a nonempty class of wormhole configurations for which both NEC$_1$ and NEC$_2$ hold simultaneously at the throat, so that the full NEC can be satisfied by the physical matter. At the critical value $\alpha l^2=-r_0^2$, NEC$_1$ is saturated while NEC$_2$ is automatically satisfied for every $b'(r_0)<1$ and finite $\Phi'(r_0)$.

\subsection{Weak Energy Condition (WEC)}

The WEC requires a non-negative energy density together with the NEC. From Eq.~\eqref{ec:rho_throat}, the energy density at the throat can be written as
\begin{equation}
	\kappa\rho\Big|_{r_0}=\frac{3}{2r_0^2}\left[2- \left(1+\frac{\alpha l^2}{r_0^2}\right) \big(1-b'(r_0)\bigr)\right].
	\label{ec:WEC_rho2}
\end{equation}
In the regime $\alpha l^2<-r_0^2$, the factor $(1+\alpha l^2/r_0^2)$ is negative, while the flare-out condition implies $1-b'(r_0)>0$. Therefore, the second term inside the square brackets in Eq.~\eqref{ec:WEC_rho2} is positive after accounting for the overall minus sign. It follows that
\begin{equation}
	\rho(r_0)>0
\end{equation}

In this regime, the positivity of the energy density and NEC$_1$ are automatically satisfied at the throat. Hence, the full WEC holds whenever NEC$_2$ is also satisfied. At $\alpha l^2=-r_0^2$, one has $\kappa\rho(r_0)=3/r_0^2$, while NEC$_1$ is saturated and NEC$_2$ is positive. Thus, the WEC is automatically satisfied at the critical value.

\subsection{Strong Energy Condition (SEC)}

In five-dimensional spacetimes, the SEC requires the NEC together with $2\rho+p_r+3p_l\geq0$. Using Eqs.~\eqref{eq:emtem}, this combination evaluated at the throat is
\begin{equation}
	\kappa\left(2\rho+p_r+3p_l\right)\Big|_{r_0}=\frac{3\big(1-b'(r_0)\big)}{2r_0^4}
	\Big(r_0\left(r_0^2+\alpha l^2\right)\Phi'(r_0)-2\alpha l^2	\Big).
	\label{ec:SEC}
\end{equation}
Since the flare-out condition implies $b'(r_0)<1$, the prefactor in Eq.~\eqref{ec:SEC} is positive. Therefore, the quantity $2\rho+p_r+3p_l$ is non-negative at the throat provided that
\begin{equation}
	\left(r_0^2+\alpha l^2\right)\Phi'(r_0)
	\geq
	\frac{2\alpha l^2}{r_0}.
	\label{ec:SEC-redshift-condition01}
\end{equation}

Since NEC$_1$ is a necessary component of the SEC, its satisfaction requires $r_0^2+\alpha l^2\leq0$. If $r_0^2+\alpha l^2=0$, Eq.~\eqref{ec:SEC-redshift-condition01} is automatically satisfied. Otherwise, this factor is negative, so dividing by it reverses the inequality, yielding
\begin{equation}
	\Phi'(r_0)
	\leq
	\frac{2\alpha l^2}
	{r_0\left(r_0^2+\alpha l^2\right)}.
	\label{ec:SEC-redshift-condition}
\end{equation}

Because $\alpha l^2<-r_0^2$ implies $\alpha<0$, both the numerator and the denominator on the right-hand side of Eq.~\eqref{ec:SEC-redshift-condition} are negative. Their ratio is therefore positive. Consequently, the SEC bound includes all configurations with $\Phi'(r_0)\leq0$ and also permits positive values of $\Phi'(r_0)$ satisfying Eq.~\eqref{ec:SEC-redshift-condition}.

Hence, the condition $\alpha l^2<-r_0^2$ guarantees NEC$_1$ at the throat but is not sufficient to ensure the full SEC. In addition, NEC$_2$ and the bound \eqref{ec:SEC-redshift-condition} must be satisfied. At $\alpha l^2=-r_0^2$, both NEC and the remaining SEC inequality are automatically satisfied.

\subsection{Dominant Energy Condition (DEC)}

The DEC requires $\rho \geq 0$ and $\rho-|p_r|\geq 0$ and $\rho - |p_l| \geq 0$. Since $p_r$ is negative at the throat, the condition $\rho-|p_r|\geq 0$ (DEC$_1$) becomes
\begin{equation}
	\kappa(\rho - |p_r|)\Big|_{r_0} = \kappa(\rho + p_r)\Big|_{r_0} =	\frac{3}
	{2r_{0}^{2}}\big(-1+b'(r_{0})\big)\left(1+\frac{\alpha l^{2}}{r_{0}^{2}}\right).
\end{equation}
This expression is identical to the NEC$_1$ combination and is therefore positive in the regime $\alpha l^2<-r_0^2$.

The lateral condition, denoted by DEC$_2$, requires $\rho - |p_l|\geq0$. If $p_l(r_0)\leq0$, this condition reduces to $\rho+p_l\geq0$, and is therefore identical to NEC$_2$. If $p_l(r_0)>0$, DEC$_2$ instead requires
\begin{equation}
	\kappa(\rho-p_l)\Big|_{r_0}=\frac{1}{2r_0^2}\left\{3+5b'(r_0)+\big(1-b'(r_0)\big) \left[-\frac{3\alpha l^2}{r_0^2}-r_0\Phi'(r_0)\left(	1+\frac{\alpha l^2}{r_0^2}	\right)	\right]	\right\}.
	\label{ec:DEC}
\end{equation}

Thus, in the regime $\alpha l^2<-r_0^2$, the positivity of $\rho$ and the radial DEC are automatically satisfied at the throat. The full DEC additionally requires the lateral condition. For $p_l(r_0)\leq0$, this is equivalent to NEC$_2$; for $p_l(r_0)>0$, Eq.~\eqref{ec:DEC} must also be non-negative. At $\alpha l^2=-r_0^2$, the lateral DEC reduces to $b'(r_0)\geq-3$. Consequently, all components of the DEC are satisfied at the critical value whenever
\begin{equation}
	-3\leq b'(r_0)<1.
\end{equation}

\subsection{Summary}

The results of this analysis are summarized in Table~\ref{tab:energy-conditions}. The key finding is that the parameter $\alpha$ plays a decisive role. In the Einstein limit, $\alpha=0$, NEC$_1$ is necessarily violated at the throat and, consequently, the full WEC, SEC, and DEC cannot be satisfied. By contrast, in the regime $\alpha l^2<-r_0^2$, the positivity of the energy density, NEC$_1$, and the radial DEC are automatically ensured. Moreover, it is possible to construct wormhole solutions for which the physical matter satisfies the full NEC, WEC, SEC, and DEC, provided that the additional constraints involving $b'(r_0)$, $\Phi'(r_0)$, and the lateral pressure are fulfilled. The critical value $\alpha l^2=-r_0^2$ is particularly favorable: the NEC, WEC, and SEC are automatically satisfied, while the DEC additionally requires $b'(r_0)\geq-3$. Thus, the EChS curvature correction allows the physical matter supporting the wormhole to satisfy the standard energy conditions at the throat for suitable choices of the metric functions and coupling parameters.

\begin{table}[h!]
	\centering
	\caption{Energy conditions evaluated at the wormhole throat for different
		regimes of the EChS coupling. ``Conditional'' indicates dependence on the metric functions or additional
		parameter restrictions.}
	\label{tab:energy-conditions}
	\renewcommand{\arraystretch}{1.25}
	\setlength{\tabcolsep}{5pt}
	\begin{tabular}{@{}lcccc@{}}
		\toprule
		Energy condition&Einstein limit $\alpha=0$&$\alpha l^{2}>-r_{0}^{2}$&
		$\alpha l^{2}=-r_{0}^{2}$&$\alpha l^{2}<-r_{0}^{2}$\\
		\midrule
		Radial NEC: $\rho+p_r\geq0$&$\times$&$\times$&$\checkmark$&$\checkmark$\\
		Lateral NEC: $\rho+p_l\geq0$&Conditional&Conditional&$\checkmark$&Conditional\\
		Positive energy density: $\rho\geq0$&Conditional&Conditional&$\checkmark$&$\checkmark$\\
		WEC&$\times$&$\times$&$\checkmark$&Conditional\\
		SEC&$\times$&$\times$&$\checkmark$&Conditional\\
		Radial DEC:	$\rho\geq|p_r|$&$\times$&$\times$&$\checkmark$&$\checkmark$\\
		Lateral DEC: $\rho\geq|p_l|$&Conditional&Conditional&Conditional&Conditional\\
		\bottomrule
	\end{tabular}
\end{table}

\subsection{On the sign of the radial pressure}

From Eq.~(\ref{ec:pr_throat}), the radial pressure at the throat satisfies
\begin{eqnarray}
	\kappa p_r = -\frac{3}{r_0^2} < 0,
\end{eqnarray}
which is strictly negative, independently of the shape function $b(r)$, the redshift function $\Phi(r)$, and the higher-curvature parameter $\alpha$. This result is purely geometric: it follows directly from the field equations when a wormhole geometry is considered, and cannot be avoided by any choice of matter content or modified gravity parameters.

This behavior stands in sharp contrast with what is observed in cosmological contexts. In a homogeneous and isotropic universe described by the Friedmann--Lema\^itre--Robertson--Walker (FLRW) metric, the effective radial and lateral pressures are isotropic by symmetry, $p_r = p_l = p$, and their sign determines the expansion history of the universe. In particular:
\begin{itemize}
	\item Ordinary matter (dust, radiation) satisfies $p \geq 0$, leading to decelerated expansion.
	\item Dark energy, modeled as a cosmological constant $\Lambda$, has effective pressure $p_\Lambda = -\rho_\Lambda < 0$, driving accelerated expansion. In this case the negative pressure is isotropic and uniform across all spatial directions.
	\item Quintessence fields admit $-\rho < p < -\rho/3$, also with negative 
	pressure but evolving in time.
\end{itemize}

The crucial difference is the following. In cosmology, a negative isotropic pressure $p < 0$ with $\rho > 0$ is perfectly compatible with the NEC and WEC, and may also satisfy the DEC: the NEC requires $\rho + p \geq 0$, which can hold even when $p < 0$, as long as $|p| \leq \rho$. A cosmological constant saturates this bound with $p = -\rho$, although it violates the SEC.

In the wormhole case, however, the situation is fundamentally different. The radial pressure is negative, $p_r < 0$, but it acts along the radial direction threading the wormhole throat, which is precisely the direction along which the NEC is most stringently tested. The NEC combination $\rho + p_r$ involves this negative radial pressure directly, and in standard general relativity ($\alpha = 0$) one has
\begin{eqnarray}
	\kappa(\rho + p_r)\big|_{\alpha=0} = \frac{3\left(-1+b'(r_0)\right)}{2r_0^2} < 0,
\end{eqnarray}
since the flare-out condition forces $b'(r_0) < 1$. Thus, in Einstein gravity, the negative radial pressure inevitably drives the NEC into violation --- a result with no cosmological analogue, since in FLRW spacetimes the accelerated-expansion regime with $p < 0$ does \emph{not} require NEC violation.

The higher-curvature correction $\alpha$ modifies this picture precisely through its contribution to $\rho + p_r$, as seen in 
Eq.~(\ref{ec:NEC1}). For $\alpha l^2 < -r_0^2 < 0$, the factor $(r_0^2 + \alpha l^2)$ becomes negative, and since $(-1 + b'(r_0)) < 0$ as well, their product is positive, allowing the NEC to be satisfied. In this sense, $\alpha$ plays a role analogous to dark energy in cosmology: it introduces an effective negative contribution to the geometry that compensates the 
destabilizing effect of the wormhole topology on the energy conditions. However, unlike the cosmological case where $\Lambda$ acts uniformly on all scales, here the correction $\alpha$ is tied to the local curvature at the throat and vanishes in the weak-field limit.

In summary, the negativity of $p_r$ at the wormhole throat is a geometric inevitability, not a pathology of the matter model. Its physical meaning is that the matter threading the throat must exert a \emph{tension} along the radial direction in order to sustain the wormhole geometry against collapse ---a role conceptually similar to the negative pressure associated with accelerated cosmological expansion, but geometrically distinct and subject to more stringent energy-condition constraints.

\section{Power-law wormhole family}\label{sec:powerlaw} 

The analytical results of the preceding section establish general conditions on the EChS coupling under which the energy conditions can be satisfied at the wormhole throat, without specifying a particular shape function. To make these results explicit and suitable for graphical analysis, we now consider the one-parameter family of shape functions given by the power-law ansatz
\begin{equation}
	b(r) = r_0 \left(\frac{r_0}{r}\right)^n, \qquad n > 0,
	\label{ec:powerlaw}
\end{equation}
where $r_0 > 0$ is the throat radius and $n$ is a free parameter controlling the rate at which the wormhole geometry opens up away from the throat. This family satisfies the standard requirements on the shape function: $b(r_0) = r_0$ (throat condition), $b(r)/r \to 0$ as $r \to \infty$ (asymptotic-flatness condition), and $b'(r_0) = -n < 0 < 1$ (flare-out condition), the latter being satisfied for all $n > 0$. Substituting $b'(r_0)=-n$ into Eq.~\eqref{ec:NEC1}, the radial NEC combination becomes
\begin{equation}
	\kappa(\rho+p_r)\Big|_{r_0}=-\frac{3(1+n)}{2r_0^2}
	\left(
	1+\frac{\alpha l^2}{r_0^2}
	\right).
	\label{ec:powerlaw-NEC1}
\end{equation}
Since $1+n>0$, the sign of Eq.~\eqref{ec:powerlaw-NEC1} is determined entirely by the factor $1+\alpha l^2/r_0^2$. Therefore, NEC$_1$ is satisfied at the throat if and only if $\alpha l^2\leq-r_0^2$, being saturated at $\alpha l^2=-r_0^2$ and strictly satisfied for $\alpha l^2<-r_0^2$. This result is independent of the power-law index $n$ and agrees with the general analysis. Similarly, the energy density at the throat \eqref{ec:WEC_rho2} becomes
\begin{equation}
	\kappa\rho\Big|_{r_0}=\frac{3}{2r_0^2}\left[2-(1+n)\left(1+\frac{\alpha l^2}{r_0^2}
	\right)\right].\label{ec:powerlaw-rho}
\end{equation}
For $\alpha l^2<-r_0^2$, the second term inside the square brackets is positive after accounting for the overall minus sign. Thus, $\rho(r_0)>0$ for every $n>0$. For the zero tidal-force case $\Phi(r) = 0$, the lateral pressure \eqref{ec:pl_throat} reduces to $\kappa p_l\big|_{r_0}=n/r_0^2$. Hence, the lateral NEC combination at the throat is then
\begin{equation}
	\kappa(\rho+p_l)\Big|_{r_0}=\frac{1}{2r_0^2}\left[3-n-3(1+n)\frac{\alpha l^2}{r_0^2}\right].
	\label{ec:powerlaw-NEC2}
\end{equation}
For $\alpha l^2<-r_0^2$, the right-hand side is strictly positive for every $n>0$, and therefore NEC$_2$ is satisfied. Moreover, in the same regime ($\Phi(r)=0$) the SEC trace combination becomes
\begin{equation}
	\kappa\left(2\rho+p_r+3p_l\right)=-\frac{3\alpha l^2(1+n)b(r)^2}{r^6}.
	\label{ec:powerlaw-SEC}
\end{equation}
Thus, when $\alpha<0$, the SEC trace combination is positive not only at the throat, but throughout the wormhole exterior. Furthermore, the radial part of the DEC is identical to NEC$_1$, since $p_r(r_0)<0$, and is therefore satisfied when $\alpha l^2\leq-r_0^2$. Because $p_l(r_0)>0$, the lateral part of the DEC requires $\rho(r_0)-p_l(r_0)\geq0$, which for the power-law family is
\begin{equation}
	\kappa(\rho-p_l)\Big|_{r_0}=\frac{1}{2r_0^2}\left(3-5n-3(1+n)\frac{\alpha l^2}{r_0^2}\right).
	\label{ec:powerlaw-DEC2}
\end{equation}
Therefore, the lateral DEC is satisfied provided that
\begin{equation}
	-\frac{\alpha l^2}{r_0^2}
	\geq
	\frac{5n-3}{3(1+n)}.
	\label{ec:powerlaw-DEC-condition}
\end{equation}
For $0<n\leq3$, the condition $\alpha l^2<-r_0^2$ is already sufficient. For $n>3$, however, the lateral DEC imposes the stronger restriction \eqref{ec:powerlaw-DEC-condition}. In summary, for the power-law wormhole family and $\alpha l^2<-r_0^2$, NEC$_1$ and positivity of the energy density are simultaneously satisfied at the throat. If, in addition, the zero tidal-force case is considered, the full NEC, WEC, SEC, and the radial part of the DEC are satisfied for every $n>0$. DEC$_2$ is also satisfied for $0<n\leq3$, whereas for $n>3$ it requires a more negative value of the EChS coupling according to Eq.~\eqref{ec:powerlaw-DEC-condition}. Therefore, for $\Phi = 0$ and a sufficiently negative value of $\alpha l^2$, all four standard energy conditions are simultaneously satisfied at the throat for every member of the power-law family~(\ref{ec:powerlaw}), regardless of the value of $n$.

\subsection{Graphical analysis}

The analytical results derived for the power-law family are illustrated in the following figures, which provide a complementary visual picture of the energy conditions along the wormhole exterior and in the parameter space $(n,\alpha l^2/r_0^2)$.

\begin{figure}[ht]
	\centering
	\includegraphics[scale=1]{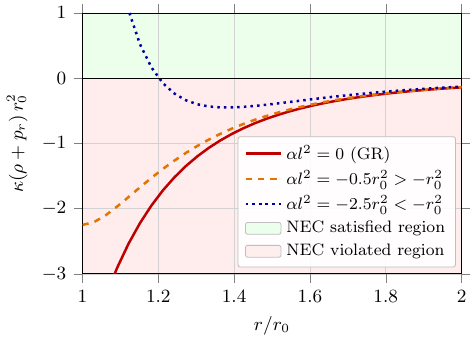}
	\caption{Null Energy Condition combination $\kappa(\rho + p_r)$ as a function of $r/r_0$ for $b(r) = r_0(r_0/r)^2$ and $\Phi = 0$, for three representative values of $\alpha l^2$. For $\alpha l^2 = 0$ and $\alpha l^2>-r_0^2$, the NEC is violated at the throat. For $\alpha l^2<-r_0^2$, the NEC is satisfied at the throat and in a finite region outside it, before becoming violated at larger $r$. Green (red) shading indicates $\kappa(\rho + p_r)>0$ ($<0$).}
	\label{fig:NEC}
\end{figure}

Figure~\ref{fig:NEC} shows the NEC combination $\kappa(\rho + p_r)$ as a function of $r/r_0$ for the representative case $n = 2$ (i.e., $b(r) = r_0(r_0/r)^2$) and $\Phi = 0$, for three values of $\alpha l^2$. For $\alpha l^2 = 0$ and $\alpha l^2 = -0.5r_0^2>-r_0^2$, the combination is negative already at the throat, confirming the NEC violation expected in these regimes. For $\alpha l^2 = -2.5r_0^2<-r_0^2$, the NEC is satisfied at the throat and in a finite region immediately outside it, but becomes negative at larger $r$. This behavior follows from the decreasing relative importance of the EChS correction away from the throat. Thus, the threshold $\alpha l^2=-r_0^2$ determines the sign of NEC$_1$ at the throat, as established analytically, while its behavior throughout the exterior also depends on $r$.

\begin{figure*}[ht]
	\centering
	\begin{subfigure}[t]{0.48\textwidth}
		\centering
		\includegraphics[scale=1]{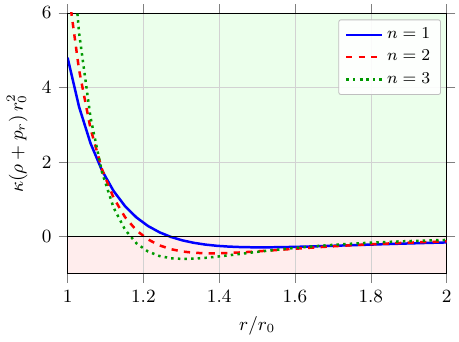}
		\caption{NEC$_1$: $\kappa(\rho+p_r)r_0^2$}
		\label{fig:panel-a}
	\end{subfigure}
	\hfill
	\begin{subfigure}[t]{0.48\textwidth}
		\centering
		\includegraphics[scale=1]{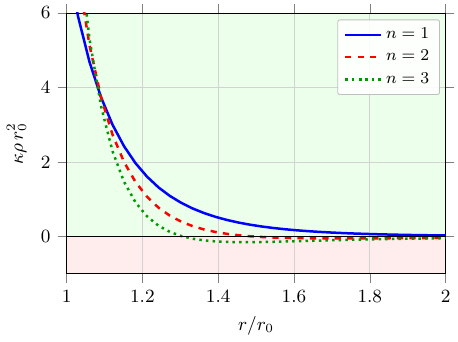}
		\caption{WEC: $\kappa\rho r_0^2$}
		\label{fig:panel-b}
	\end{subfigure}
	
	\vspace{0.5cm}
	
	\begin{subfigure}[t]{0.48\textwidth}
		\centering
		\includegraphics[scale=1]{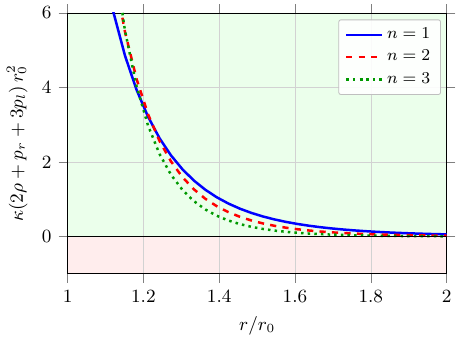}
		\caption{SEC: $\kappa(2\rho+p_r+3p_l)r_0^2$}
		\label{fig:panel-c}
	\end{subfigure}
	\hfill
	\begin{subfigure}[t]{0.48\textwidth}
		\centering
		\includegraphics[scale=1]{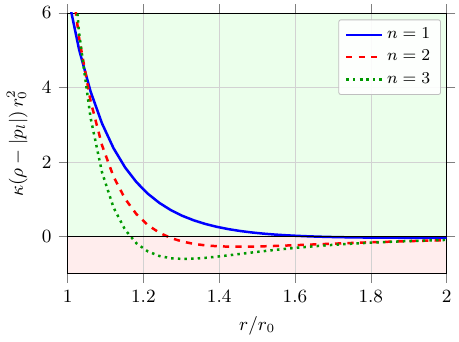}
		\caption{DEC$_2$: $\kappa(\rho-|p_l|)r_0^2$}
		\label{fig:panel-d}
	\end{subfigure}
	
	\caption{Energy-condition combinations along the wormhole exterior for $b(r)=r_0(r_0/r)^n$, $\Phi=0$, and $\alpha l^2=-2.5r_0^2$, for $n=1$ (solid), $n=2$ (dashed), and $n=3$ (dotted). Top left: NEC$_1$, $\kappa(\rho+p_r)$. Top right: energy density $\kappa\rho$. Bottom left: SEC trace combination $\kappa(2\rho+p_r+3p_l)$. Bottom right: DEC$_2$, $\kappa(\rho-|p_l|)$. The relevant combinations are positive at the throat for the three values of $n$ shown, although NEC$_1$, the energy density, and DEC$_2$ may change sign farther from the throat.}
	\label{fig:all_EC}
\end{figure*}

Figure~\ref{fig:all_EC} presents the relevant energy-condition combinations for $\alpha l^2=-2.5r_0^2$ and three values of the power-law index, $n=1,2,3$, with $\Phi=0$. The NEC$_1$ combination $\kappa(\rho+p_r)$ (top left) is positive at the throat for all three values of $n$, but becomes negative farther from the throat. The energy density $\kappa\rho$ (top right) is also positive at the throat, although for the steeper profiles it changes sign at larger $r$. The SEC trace combination $\kappa(2\rho+p_r+3p_l)$ (bottom left) remains positive throughout the plotted domain, consistently with Eq.~\eqref{ec:powerlaw-SEC}. Finally, the DEC$_2$ combination $\kappa(\rho-|p_l|)$ (bottom right) is positive at the throat for the three values shown but may become negative farther away. The figure therefore confirms that the simultaneous satisfaction of the energy conditions at the throat is robust for the displayed members of the power-law family, while this property does not in general extend throughout the wormhole exterior.

\begin{figure*}[htbp]
	\centering
	\includegraphics[scale=1]{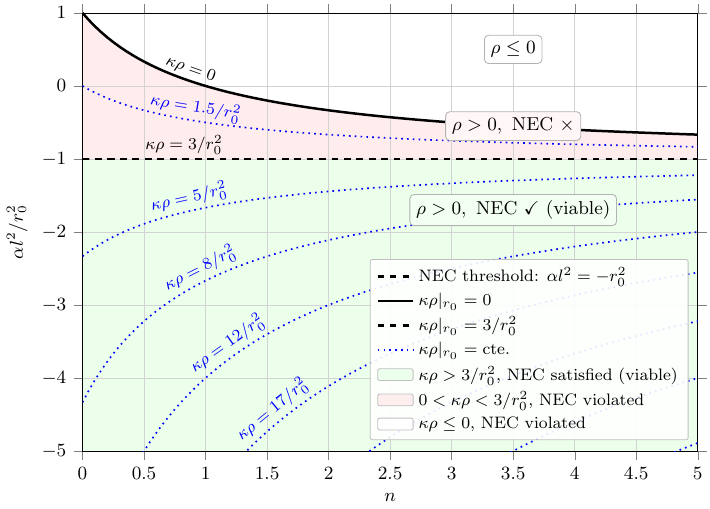}
	\caption{Parameter space $(n,\alpha l^2/r_0^2)$ for the power-law shape function $b(r)=r_0(r_0/r)^n$, showing the sign of the energy density $\kappa\rho$ and the status of the Null Energy Condition (NEC) at the wormhole throat. The dashed horizontal line marks the NEC threshold $\alpha l^2=-r_0^2$, independent of $n$. The solid curve marks the boundary $\kappa\rho\big|{r_0}=0$, given by $\alpha l^2=r_0^2(1-n)/(1+n)$, which crosses $\alpha=0$ at $n=1$. The viable region below the NEC threshold has both $\rho>0$ and the NEC satisfied at the throat. The intermediate region has $\rho>0$ with NEC violated, while the region above the solid curve has $\rho\leq0$. Dotted curves show representative contour levels of $\kappa\rho\big|{r_0}$.}
	\label{fig:param_space_rho}
\end{figure*}

Figure~\ref{fig:param_space_rho} displays the parameter space $(n,\alpha l^2/r_0^2)$ and identifies the regions where the energy density is positive and the NEC is simultaneously satisfied at the throat. The plane is divided by two relevant boundaries. The first is the horizontal line $\alpha l^2=-r_0^2$, which constitutes the NEC threshold: for $\alpha l^2<-r_0^2$ the NEC is satisfied at the throat for all $n>0$, while for $\alpha l^2>-r_0^2$ it is violated regardless of $n$. The second boundary is the curve
\begin{equation}
	\alpha l^2=r_0^2\frac{1-n}{1+n},
\end{equation}
which marks the locus $\kappa\rho\big|_{r_0}=0$ and intersects $\alpha=0$ at $n=1$.

One can verify analytically that whenever $\alpha l^2<-r_0^2$, the energy density at the throat is strictly positive. Indeed,
\begin{equation}
	\kappa\rho\Big|_{r_0}=\frac{3}{2r_0^4}\left[r_0^2(1-n)-\alpha l^2(1+n)\right],
\end{equation}
and since $\alpha l^2<-r_0^2$, one has $-\alpha l^2(1+n)>r_0^2(1+n)$. Consequently,
\begin{equation}
	r_0^2(1-n)-\alpha l^2(1+n)>2r_0^2>0.
\end{equation}
Thus, no point satisfying NEC$_1$ at the throat can have negative energy density there.

The parameter space therefore contains three distinct sectors. The region $\alpha l^2<-r_0^2$ is the physically viable sector where both $\rho>0$ and the NEC hold simultaneously at the throat. Above the NEC threshold but below the $\kappa\rho=0$ curve, the energy density remains positive while the NEC is violated, so the matter source remains exotic in the usual sense. Above the $\kappa\rho=0$ curve, the energy density is non-positive and the NEC is also violated. In all cases, the criterion $\alpha l^2<-r_0^2$ provides a sharp, $n$-independent condition for positive energy density together with NEC satisfaction at the wormhole throat.

\subsection{Energy conditions at the throat as functions of $\alpha l^2$} 
\begin{figure*}[htbp]
	\centering
	\includegraphics[scale=1]{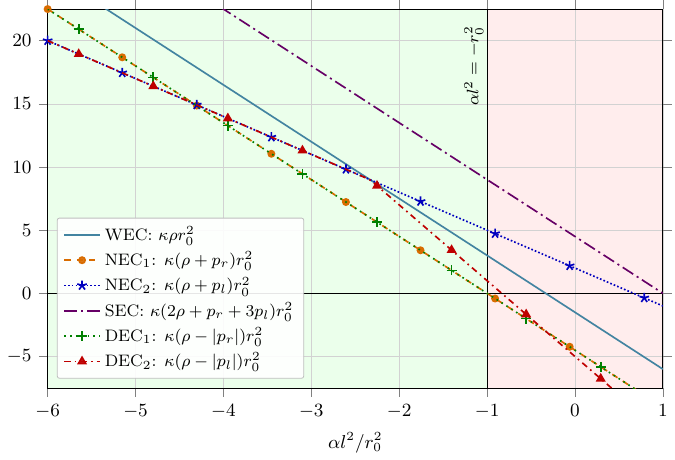}
	\caption{Energy-condition combinations at the wormhole throat as functions of $\alpha l^2/r_0^2$, for $b'(r_0)=-2$ and $\Phi'(r_0)=1/r_0$ (power-law $n=2$). The curves correspond to the energy density $\kappa\rho$, NEC$_1$ $\kappa(\rho+p_r)$, NEC$_2$ $\kappa(\rho+p_l)$, SEC $\kappa(2\rho+p_r+3p_l)$, DEC$_1$ $\kappa(\rho-|p_r|)$, and DEC$_2$ $\kappa(\rho-|p_l|)$. The black dotted vertical line marks the NEC threshold $\alpha l^2=-r_0^2$. Green (red) shading indicates the NEC-satisfying (NEC-violating) regime. All conditions are satisfied simultaneously for $\alpha l^2\leq-r_0^2$.}
	\label{fig:EC_throat}
\end{figure*}

The analytical expressions~(\ref{ec:rho_throat})--(\ref{ec:pl_throat}) allow one to study how each energy condition at the throat varies continuously with the EChS coupling $\alpha l^2$, for fixed values of the geometric parameters $b'(r_0)$ and $\Phi'(r_0)$. This provides a complementary perspective to the analysis of Section~\ref{sec:EC}: rather than fixing $\alpha l^2$ and varying the shape function, we fix the shape function and trace the transition of each energy condition as $\alpha l^2$ decreases from zero (GR limit) into the non-exotic regime. Figure~\ref{fig:EC_throat} shows all six standard energy-condition combinations at the throat as functions of $\alpha l^2/r_0^2$, for the representative parameter choice $b'(r_0)=-2$ (corresponding to the power-law index $n=2$), $\Phi'(r_0)=1/r_0$, and $r_0=1$. The black dotted vertical line marks the NEC threshold $\alpha l^2=-r_0^2$. Green (red) shading indicates the NEC-satisfying (NEC-violating) regime. The throat quantities~(\ref{ec:rho_throat})--(\ref{ec:pl_throat}) are linear in $\alpha l^2$, so all combinations are linear except DEC$_2$, which is piecewise linear because of the absolute value $|p_l|$.

The results are the following. The energy density $\kappa\rho$ becomes positive at $\alpha l^2=-r_0^2/3$, which lies within the NEC-violating regime: positive energy density is therefore reached before the radial NEC threshold is crossed. NEC$_1$, $\kappa(\rho+p_r)$, changes sign exactly at $\alpha l^2=-r_0^2$, confirming the threshold derived in~(\ref{ec:NEC1}). NEC$_2$, $\kappa(\rho+p_l)$, is positive throughout the NEC-satisfying regime and changes sign only at $\alpha l^2=2r_0^2/3$. The SEC combination $\kappa(2\rho+p_r+3p_l)$ is likewise strictly positive throughout the NEC-satisfying regime and changes sign only at $\alpha l^2=r_0^2$. DEC$_1$, $\kappa(\rho-|p_r|)$, coincides identically with NEC$_1$, since $p_r<0$ always, and therefore changes sign at $\alpha l^2=-r_0^2$. Finally, DEC$_2$, $\kappa(\rho-|p_l|)$, changes sign at $\alpha l^2=-5r_0^2/6$. Since $p_l$ changes sign at $\alpha l^2=-7r_0^2/3$, DEC$_2$ is piecewise linear, but remains positive throughout the NEC-satisfying regime $\alpha l^2\leq-r_0^2$.

Three features of this analysis deserve emphasis. First, the relevant thresholds encountered as $\alpha l^2$ decreases from the GR limit are

\begin{equation}
	\alpha l^2_{\rm WEC}=-\frac{r_0^2}{3}
	>
	\alpha l^2_{\rm DEC_2}=-\frac{5r_0^2}{6}
	>
	-r_0^2=
	\alpha l^2_{\rm NEC_1}=
	\alpha l^2_{\rm DEC_1}.
	\label{ec:threshold_ordering}
\end{equation}

Thus, positive energy density and DEC$_2$ are reached before the radial NEC threshold, while NEC$_1$, NEC$_2$, WEC, SEC, DEC$_1$, and DEC$_2$ are all satisfied simultaneously for $\alpha l^2\leq-r_0^2$. Second, this confirms the main result of Section~\ref{sec:EC} for this specific parameter choice and shows that the critical value $\alpha l^2=-r_0^2$ already suffices, with NEC$_1$ and DEC$_1$ saturated there. Third, the slopes and zero-crossings are determined by the coefficients of $\alpha l^2$ in~(\ref{ec:rho_throat})--(\ref{ec:pl_throat}) and depend on $b'(r_0)$ and $r_0\Phi'(r_0)$. Different choices of these geometric quantities shift the individual thresholds, while the universal radial NEC threshold $\alpha l^2=-r_0^2$ remains unchanged.

\subsection{Throat values as initial conditions and the isotropic pressure case}\label{sec:throat_IC} The expressions~(\ref{ec:rho_throat})--(\ref{ec:pl_throat}) are not merely useful for analyzing energy conditions at a single point: they provide the matter variables at the throat in terms of the local geometric data. Given the throat radius $r_0$, the higher-curvature parameter $\alpha$, and a choice of $b'(r_0)$ and $\Phi'(r_0)$ satisfying the flare-out condition $b'(r_0)<1$, the values of $\rho$, $p_r$, and $p_l$ at $r=r_0$ are fully determined by~(\ref{ec:rho_throat})--(\ref{ec:pl_throat}). These quantities, together with the geometric condition $b(r_0)=r_0$ and a choice of $\Phi(r_0)$ (which can always be set to zero by a rescaling of the time coordinate), provide natural throat data for a numerical integration toward the asymptotic region $r\to\infty$, once an equation of state or another closure condition is supplied. In this sense, the throat provides a natural starting point for the radial integration, and the parameters $(\alpha,b'(r_0),\Phi'(r_0))$ characterize the local geometric data of the problem. 

A particularly interesting specialization arises when one imposes \emph{isotropic pressure} at the throat, $p_r(r_0)=p_l(r_0)$. Equating~(\ref{ec:pr_throat}) and~(\ref{ec:pl_throat}) yields the algebraic condition
\begin{equation}
	\left(b'(r_0)-1\right)\Phi'(r_0)\left(r_0^2+\alpha l^2\right)
	+2r_0\left(b'(r_0)-3\right)=0.
	\label{ec:isotropic_condition}
\end{equation}
This can be rewritten as
\begin{equation}
	\left(b'(r_0)-1\right)\Phi'(r_0)\left(r_0^2+\alpha l^2\right)
	=2r_0\left(3-b'(r_0)\right),
	\label{ec:isotropic_condition2}
\end{equation}
where both sides depend on $b'(r_0)$. Since the flare-out condition requires $b'(r_0)<1$, the left-hand side has the sign of $-\Phi'(r_0)(r_0^2+\alpha l^2)$, while the right-hand side is positive because $b'(r_0)<3$ follows automatically from the flare-out condition. From~(\ref{ec:isotropic_condition}), the derivative of the shape function at the throat is determined by $\Phi'(r_0)$ and $\alpha$:
\begin{equation}
	b'(r_0)=\frac{\Phi'(r_0)\left(r_0^2+\alpha l^2\right)+6r_0}
	{\Phi'(r_0)\left(r_0^2+\alpha l^2\right)+2r_0}.
	\label{ec:bp_isotropic}
\end{equation}
This is the key consequence of isotropic pressure at the throat: the free local data are reduced from three independent parameters $(\alpha,b'(r_0),\Phi'(r_0))$ to two, namely $(\alpha,\Phi'(r_0))$, with $b'(r_0)$ fixed by~(\ref{ec:bp_isotropic}). Indeed,
\begin{equation}
	b'(r_0)-1=
	\frac{4r_0}{\Phi'(r_0)\left(r_0^2+\alpha l^2\right)+2r_0},
\end{equation}
so the flare-out condition $b'(r_0)<1$ is equivalent to
\begin{equation}
	\Phi'(r_0)\left(r_0^2+\alpha l^2\right)<-2r_0.
	\label{ec:isotropic_flareout}
\end{equation}
Substituting~(\ref{ec:bp_isotropic}) into~(\ref{ec:rho_throat}), the energy density at the throat under isotropic pressure takes the form
\begin{equation}
	\kappa\rho\big|_{r_0}=
	\frac{3\left[2\alpha l^2+r_0\left(\Phi'(r_0)(r_0^2+\alpha l^2)+4r_0\right)\right]}
	{r_0^3\left[\Phi'(r_0)(r_0^2+\alpha l^2)+2r_0\right]},
	\label{ec:rho_isotropic}
\end{equation}
which retains explicit dependence on $\alpha l^2$. Similarly, substituting~(\ref{ec:bp_isotropic}) into~(\ref{ec:NEC1}) yields
\begin{equation}
	\kappa(\rho+p_r)\big|_{r_0}=
	\frac{6\left(r_0^2+\alpha l^2\right)}
	{r_0^3\left[\Phi'(r_0)\left(r_0^2+\alpha l^2\right)+2r_0\right]}.
	\label{ec:NEC_isotropic}
\end{equation}
Under the flare-out condition \eqref{ec:isotropic_flareout}, the denominator in~(\ref{ec:NEC_isotropic}) is negative. Therefore, the NEC at the throat is satisfied if $r_0^2+\alpha l^2<0$, that is, if $\alpha l^2<-r_0^2$. The limiting value $\alpha l^2=-r_0^2$ cannot describe an isotropic wormhole throat, since Eq.~\eqref{ec:bp_isotropic} then gives $b'(r_0)=3$, in direct conflict with the flare-out condition. Thus, in the isotropic case, the non-exotic regime requires the strict inequality $\alpha l^2<-r_0^2$, consistently with the threshold identified in the anisotropic analysis. 

Two features of this analysis deserve emphasis. First, Eqs.~(\ref{ec:rho_isotropic}) and~(\ref{ec:NEC_isotropic}) follow directly from the exact throat field equations of EChS gravity, without restriction to a particular shape function. Second, imposing $p_r=p_l$ globally for all $r\geq r_0$ reduces the independent metric functions by one relation. Once $\Phi(r)$ is specified, the isotropy condition provides a first-order differential equation for $b(r)$, with the throat values satisfying~(\ref{ec:isotropic_condition}) as consistent boundary data for the radial integration. 

It is instructive to examine the particular case $\Phi'(r_0)=0$ within the isotropic framework. Setting $\Phi'(r_0)=0$ in~(\ref{ec:isotropic_condition}) immediately yields $b'(r_0)=3$, which violates the flare-out condition $b'(r_0)<1$. This can also be seen directly from~(\ref{ec:bp_isotropic}): with $\Phi'(r_0)=0$ the numerator and denominator reduce to $6r_0$ and $2r_0$, respectively, giving $b'(r_0)=3$ independently of $\alpha l^2$. Therefore, isotropic pressure at the throat is \emph{locally} incompatible with $\Phi'(r_0)=0$, regardless of the value of the EChS coupling. This local result anticipates and reinforces the global conclusion established below: no globally isotropic wormhole solution exists for $\Phi(r)=0$ within EChS gravity, since the incompatibility is already present at the throat. 

The analysis of the energy conditions in this global setting requires further investigation, since the simplifications that yield~(\ref{ec:rho_isotropic}) and~(\ref{ec:NEC_isotropic}) are specific to the throat and do not persist along the entire exterior spacetime. In the particular case $\Phi(r)=0$, the condition $p_r=p_l$ for all $r\geq r_0$ reduces to the ODE $b'=3b/r$, whose unique solution satisfying $b(r_0)=r_0$ is $b(r)=r^3/r_0^2$. This violates both the flare-out condition ($b'(r_0)=3>1$) and asymptotic flatness ($b(r)/r\to\infty$), and is therefore not a physically viable wormhole. Hence, no globally isotropic wormhole solution exists for $\Phi=0$ within EChS gravity. If a globally isotropic solution exists, it necessarily requires a non-trivial redshift function $\Phi(r)\neq0$, and its construction demands the integration of the resulting differential equation relating $\Phi(r)$ and $b(r)$, either numerically or through an appropriate ansatz.

\subsection{Volume Integral Quantifier} A useful global measure of the ``total amount'' of NEC-violating matter threading a wormhole is provided by the Volume Integral Quantifier (VIQ), introduced by Visser, Kar and Dadhich~\cite{Visser:2003yf}. For a five-dimensional wormhole with the Morris--Thorne metric~\eqref{eq:Morris-Thorne-metric}, we define the VIQ as
\begin{equation}
	\mathcal{I} = \int_{r_0}^{\infty} (\rho + p_r)\, r^3\, dr,
	\label{ec:VIQ_def}
\end{equation}
where the factor $r^3$ corresponds to the radial coordinate-volume measure associated with the three-sphere, up to the constant angular factor $2\pi^2$, which is suppressed for brevity. A value $\mathcal{I}<0$ signals a negative net integrated NEC contribution, while $\mathcal{I}\geq0$ indicates that the integrated NEC combination is non-negative. The latter is a necessary, though not sufficient, condition for pointwise absence of NEC-violating matter throughout the exterior. In GR ($\alpha=0$), the flare-out condition forces $\rho+p_r<0$ at the throat. For the power-law family considered below with $\Phi=0$, the GR contribution remains negative throughout the exterior and therefore gives $\mathcal{I}<0$.

\subsubsection*{General result} Using the NEC combination at general $r$, for the power-law shape function $b(r)=r_0(r_0/r)^n$ with $\Phi=0$, one obtains
\begin{equation}
	\kappa(\rho + p_r) = -\frac{3(n+1)}{2}\,\frac{r_0^{n+1}}{r^{n+3}}
	\left(1 + \frac{\alpha l^2\, r_0^{n+1}}{r^{n+3}}\right).
	\label{ec:NEC_integrand}
\end{equation}
This expression separates cleanly into a GR contribution and an EChS correction proportional to $\alpha l^2$. Substituting into~(\ref{ec:VIQ_def}) and integrating from $r_0$ to infinity, for $n>1$ both terms converge and yield the exact closed-form result
\begin{equation}
	\kappa\mathcal{I} = -\frac{3(n+1)}{2(n-1)}\,r_0^2
	- \frac{3}{4}\,\alpha l^2.
	\label{ec:VIQ_exact}
\end{equation}
The first term is the GR contribution, which is strictly negative for all $n>1$. The second term is the EChS correction: for $\alpha l^2<0$, it is strictly positive and reduces the magnitude of the negative integrated NEC contribution relative to GR until the VIQ vanishes.

For $n=1$, the GR term diverges logarithmically ($\mathcal{I}{\rm GR}\sim-3r_0^2\ln(R/r_0)/\kappa$ as $R\to\infty$), while the EChS correction remains finite:
\begin{equation}
	\kappa\mathcal{I}\big|_{n=1} = -3r_0^2\ln\left(\frac{R}{r_0}\right)
	- \frac{3}{4}\,\alpha l^2 + \mathcal{O}(R^{-4}),
	\label{ec:VIQ_n1}
\end{equation}
so that EChS introduces a finite positive shift over a divergent negative GR contribution.

\subsubsection*{Comparison with GR and critical coupling} Setting $\alpha l^2=0$ in~(\ref{ec:VIQ_exact}) recovers the pure GR result:
\begin{equation}
	\kappa\mathcal{I}_{\rm GR} = -\frac{3(n+1)}{2(n-1)}\,r_0^2 < 0,
	\qquad n > 1.
	\label{ec:VIQ_GR}
\end{equation}
The ratio of the magnitude of the EChS VIQ to the GR VIQ is
\begin{equation}
	\frac{|\mathcal{I}_{\rm EChS}|}{|\mathcal{I}_{\rm GR}|} =
	\left|1 + \frac{\alpha l^2\,(n-1)}{2r_0^2(n+1)}\right|.
	\label{ec:VIQ_ratio}
\end{equation}
For $\alpha l^2<0$, this ratio decreases from unity as the coupling moves away from the GR limit, reaching zero at the critical coupling. Beyond this point it increases again because $\mathcal{I}$ changes sign. Before the critical point, the reduction is more pronounced for larger values of $n$.

A particularly significant value is the \emph{critical coupling} $\alpha l^2_{\rm crit}$ at which the VIQ vanishes:
\begin{equation}
	\alpha l^2_{\rm crit} = -\frac{2r_0^2(n+1)}{n-1}.
	\label{ec:VIQ_crit}
\end{equation}
For $\alpha l^2<\alpha l^2_{\rm crit}$, the VIQ becomes positive, indicating that the positive contribution to the integrated NEC balance dominates the negative one. Numerical values of $\alpha l^2_{\rm crit}/r_0^2$ are given in Table~\ref{tab:VIQ} for representative values of $n$, together with the GR VIQ and the NEC threshold $\alpha l^2=-r_0^2$.

\begin{table}[h!]
	\centering
	\caption{GR VIQ, critical coupling, and NEC threshold for the power-law family $b(r)=r_0(r_0/r)^n$, $\Phi=0$.}
	\label{tab:VIQ}
	\begin{tabular}{cccc}
		\hline\hline
		$n$ & $\kappa\mathcal{I}_{\rm GR}/r_0^2$ &
		$\alpha l^2_{\rm crit}/r_0^2$ &
		NEC threshold $\alpha l^2/r_0^2$ \\
		\hline
		2 & $-4.500$ & $-6.000$ & $-1$ \\
		3 & $-3.000$ & $-4.000$ & $-1$ \\
		4 & $-2.500$ & $-3.333$ & $-1$ \\
		5 & $-2.250$ & $-3.000$ & $-1$ \\
		10 & $-1.833$ & $-2.444$ & $-1$ \\
		\hline\hline
	\end{tabular}
\end{table}

It is important to note that $\alpha l^2_{\rm crit}<-r_0^2$ for all $n>1$: the critical coupling lies strictly within the regime in which the NEC is satisfied at the throat. Thus, as $\alpha l^2$ decreases from zero, the radial NEC first becomes satisfied at the throat at $\alpha l^2=-r_0^2$, while the VIQ remains negative. It vanishes only at the stronger coupling $\alpha l^2=\alpha l^2_{\rm crit}$ and becomes positive beyond it. This hierarchy
\begin{equation}
	\alpha l^2_{\rm crit} < -r_0^2 < 0
	\label{ec:hierarchy}
\end{equation}
constitutes one of the central quantitative results of this analysis: NEC satisfaction at the throat and a non-negative global VIQ are compatible but distinct conditions, both achievable within EChS gravity for sufficiently negative $\alpha l^2$. A non-negative VIQ, however, does not imply pointwise NEC satisfaction throughout the exterior.

\begin{figure*}[ht]
	\centering
	\begin{subfigure}[t]{0.48\textwidth}
		\centering
		\includegraphics[scale=1]{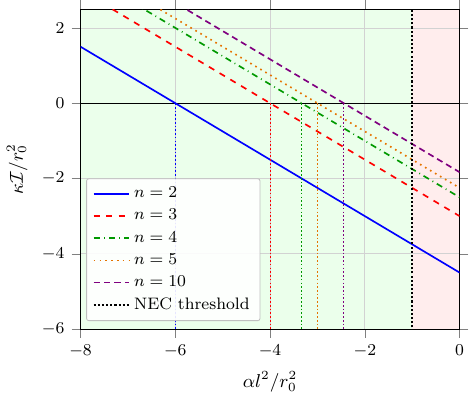}
		\caption{VIQ: $\kappa\mathcal I$ vs $\alpha$}
		\label{fig05:panel-a}
	\end{subfigure}
	\hfill
	\begin{subfigure}[t]{0.48\textwidth}
		\centering
		\includegraphics[scale=1]{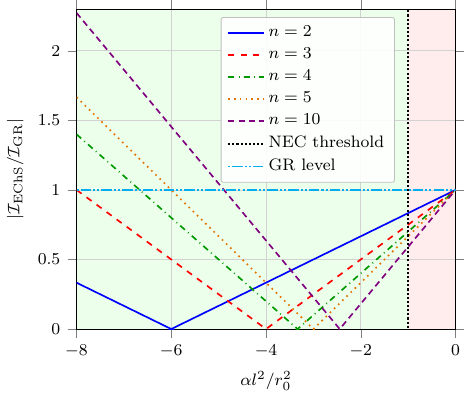}
		\caption{Magnitude of the VIQ relative to GR}
		\label{fig05:panel-b}
	\end{subfigure}
	\caption{Volume Integral Quantifier for the power-law family $b(r)=r_0(r_0/r)^n$ with $\Phi=0$, for $n=2,3,4,5,10$. Left: $\kappa\mathcal{I}/r_0^2$ as a function of $\alpha l^2/r_0^2$. The black dotted vertical line marks the NEC threshold $\alpha l^2=-r_0^2$; colored dashed lines mark the critical couplings $\alpha l^2_{\rm crit}$ at which $\mathcal{I}=0$. Green (red) shading indicates the NEC-satisfying (NEC-violating) regime at the throat. Right: ratio $|\mathcal{I}{\rm EChS}|/|\mathcal{I}{\rm GR}|$; the dashed grey line marks the GR level (ratio $=1$). At the NEC threshold, the magnitude of the VIQ is reduced relative to GR by approximately $17\%$ for $n=2$ and $41\%$ for $n=10$.}
	\label{fig:VIQ}
\end{figure*}

Figure~\ref{fig:VIQ} illustrates these results. The left panel shows $\kappa\mathcal{I}/r_0^2$ as a function of $\alpha l^2/r_0^2$ for $n=2,3,4,5,10$. All curves are linear in $\alpha l^2$, with slope $-3/4$ independent of $n$, as follows from~(\ref{ec:VIQ_exact}). The curves are vertically separated by their different GR intercepts: at $\alpha l^2=0$, each curve takes the value $\kappa\mathcal{I}_{\rm GR}/r_0^2=-3(n+1)/[2(n-1)]$, which is more negative for smaller $n$ and approaches $-3/2$ from below as $n\to\infty$. As $\alpha l^2$ decreases into the NEC-satisfying regime at the throat, all curves rise linearly, reflecting the progressive reduction of the negative integrated NEC contribution. The colored dashed vertical lines mark the critical couplings $\alpha l^2{\rm crit}$ at which $\mathcal{I}=0$. To the left of each critical line the VIQ becomes positive. All critical lines lie strictly to the left of the NEC threshold $\alpha l^2=-r_0^2$ (black dotted line), confirming the hierarchy~(\ref{ec:hierarchy}).

The right panel shows the ratio~(\ref{ec:VIQ_ratio}), $|\mathcal{I}{\rm EChS}|/|\mathcal{I}{\rm GR}|$, which measures the magnitude of the integrated NEC balance relative to GR. At $\alpha l^2=0$ all curves start at unity. As $\alpha l^2$ decreases, each curve falls toward zero at the corresponding $\alpha l^2_{\rm crit}$ and then rises again after $\mathcal{I}$ changes sign, as required by the absolute value in~(\ref{ec:VIQ_ratio}). At the NEC threshold $\alpha l^2=-r_0^2$, the magnitude of the VIQ is already reduced relative to GR by approximately $17\%$ for $n=2$ and $41\%$ for $n=10$. Thus, steeper shape functions show a larger fractional reduction of the negative integrated NEC contribution before the critical coupling is reached.

Finally, it is instructive to verify that all results reduce consistently to those of standard General Relativity in the limit $l\to0$ with $\alpha$ fixed. Since the EChS correction enters exclusively through the combination $\alpha l^2$, this limit is equivalent to $\alpha l^2\to0$. In this limit, the throat quantities~(\ref{ec:rho_throat})--(\ref{ec:pl_throat}) reduce to their GR counterparts, and in particular the NEC combination at the throat becomes
\begin{equation}
	\kappa(\rho+p_r)\Big|_{r_0,\,l\to0} =
	\frac{3\big(-1+b'(r_0)\big)}{2r_0^2} < 0,
\end{equation}
which is strictly negative by the flare-out condition $b'(r_0)<1$, recovering the standard result that exotic matter is unavoidable at the throat in GR. Correspondingly, the condition $\alpha l^2<-r_0^2$ cannot be satisfied as $l\to0$, since $\alpha l^2\to0>-r_0^2$. For the VIQ, the limit $\alpha l^2\to0$ in~(\ref{ec:VIQ_exact}) immediately gives $\kappa\mathcal{I}\to\kappa\mathcal{I}{\rm GR}=-3(n+1)r_0^2/[2(n-1)]<0$, and the ratio~(\ref{ec:VIQ_ratio}) approaches unity, confirming that the GR integrated NEC balance is recovered. For fixed $\alpha<0$, the EChS correction therefore acts as a continuous deformation of the GR result: as $|l|$ increases from zero, the negative VIQ decreases in magnitude, vanishes at $\alpha l^2=\alpha l^2_{\rm crit}$, and becomes positive beyond the critical coupling.

\section{Conclusions}\label{sec:conclusions}

In this work we have investigated traversable wormhole solutions within Einstein--Chern--Simons gravity, a five-dimensional gauge-theoretic extension of General Relativity whose field equations contain a Gauss--Bonnet-like higher-curvature correction parametrized by the combination $\alpha l^2$. The analysis was carried out directly for the physical matter sector, without absorbing any NEC violation into effective geometric contributions, providing a physically transparent treatment of the energy conditions. 

The central result is the sharp threshold $\alpha l^2=-r_0^2$: the radial NEC combination $\kappa(\rho+p_r)|{r_0}=3(-1+b'(r_0))(r_0^2+\alpha l^2)/(2r_0^4)$ is negative for $\alpha l^2>-r_0^2$, vanishes at $\alpha l^2=-r_0^2$, and becomes positive for $\alpha l^2<-r_0^2$, independently of the shape function and the redshift derivative. In the regime $\alpha l^2<-r_0^2$, the energy density is strictly positive, while the full NEC, WEC, SEC, and DEC can be simultaneously satisfied for suitable values of $b'(r_0)$ and $\Phi'(r_0)$. At the critical value $\alpha l^2=-r_0^2$, the NEC, WEC, and SEC are automatically satisfied at the throat, while the DEC imposes the additional condition $b'(r_0)\geq-3$. For the power-law family with $\Phi=0$, all four conditions can likewise be satisfied simultaneously at the throat for sufficiently negative $\alpha l^2$. 

A second exact result is the universal negativity of the radial pressure at the throat, $\kappa p_r=-3/r_0^2<0$, which is independent of $\alpha$, $b(r)$, and $\Phi(r)$. This is a geometric inevitability derived from the field equations: the matter threading the wormhole must exert a tension along the radial direction to sustain the geometry against collapse, regardless of the gravitational theory. The EChS correction does not alter this geometric fact, but it modifies the NEC combination $\rho+p_r$ through the energy density and can overcome the negative contribution of $p_r$. For the power-law family $b(r)=r_0(r_0/r)^n$ with $\Phi=0$, we derived the exact closed-form expression for the Volume Integral Quantifier
\begin{equation}
	\kappa\mathcal{I}=-\frac{3(n+1)}{2(n-1)}r_0^2-\frac{3}{4}\alpha l^2,
	\qquad n>1.
\end{equation}
The corresponding ratio to the GR value is
\begin{equation}
	\frac{|\mathcal{I}_{\rm EChS}|}{|\mathcal{I}_{\rm GR}|}
	=
	\left|1+\frac{\alpha l^2(n-1)}{2r_0^2(n+1)}\right|.
\end{equation}
Before the VIQ changes sign, the EChS correction reduces the magnitude of the negative integrated NEC contribution, with a larger fractional reduction for larger $n$. At the NEC threshold $\alpha l^2=-r_0^2$, this reduction ranges from approximately $17\%$ for $n=2$ to $41\%$ for $n=10$. A critical coupling
\begin{equation}
	\alpha l^2_{\rm crit}=-\frac{2r_0^2(n+1)}{n-1}
\end{equation}
exists at which the VIQ vanishes. The hierarchy $\alpha l^2_{\rm crit}<-r_0^2<0$ establishes that NEC satisfaction at the throat and a non-negative global VIQ are compatible but distinct conditions. A non-negative VIQ, however, does not imply pointwise NEC satisfaction throughout the wormhole exterior. 

In the limit $l\to0$ (equivalently $\alpha l^2\to0$), all results reduce continuously to those of standard GR: the radial NEC is violated at the throat and the VIQ of the power-law family is negative. The EChS correction thus acts as a continuous deformation of the GR result, allowing the physical matter at the throat to pass from an exotic to a non-exotic regime as $\alpha l^2$ crosses the critical value $-r_0^2$, while the integrated NEC balance becomes non-negative only at the stronger coupling $\alpha l^2_{\rm crit}$. 

In the isotropic pressure case, we showed analytically that imposing $p_r=p_l$ globally with $\Phi=0$ leads to the shape function $b(r)=r^3/r_0^2$, which violates both the flare-out condition and asymptotic flatness. Moreover, already at the throat, isotropy with $\Phi'(r_0)=0$ implies $b'(r_0)=3$. Therefore, no globally isotropic wormhole solution exists for $\Phi=0$ within EChS gravity; any viable isotropic solution necessarily requires a non-trivial redshift function $\Phi(r)\neq0$. 

Several directions for future work emerge naturally from this analysis. The construction of globally complete wormhole solutions satisfying the energy conditions throughout the exterior spacetime $r\geq r_0$ requires solving the field equations away from the throat with appropriate matter closure conditions and boundary data. The isotropic pressure case with $\Phi(r)\neq0$ is particularly compelling in this regard: as shown in Section~\ref{sec:throat_IC}, the condition $p_r=p_l$ provides a differential relation between $\Phi(r)$ and $b(r)$, whose integration may yield globally isotropic wormhole solutions. Since $\Phi=0$ leads to no viable solution, non-trivial redshift functions constitute a natural target for future analytical and numerical investigations.

\section*{\large Acknowledgements}

M. Cataldo was supported by the Direcci\'on de Investigaci\'on y Creaci\'on Art\'istica at the Universidad del B\'io-B\'io through grants No. RE2320220 and GI2310339. C. Quinzacara acknowledges support from ANID-Chile through FONDECYT Grant 11231238. P. Salgado was supported in part by ANID-Chile through FONDECYT Grant N° 1262414 and in part by UNAP-VRII grant N° 091/25.

\appendix

\titleformat{\section}[block]
{\normalfont\Large\bfseries\raggedright}
{Appendix~\thesection:}
{0.5em}
{}

\section{Field equations of Einstein--Chern--Simons gravity}
\label{app:EChS-field-equations}

In this appendix, we summarize the derivation of the field equations of five-dimensional Einstein--Chern--Simons gravity and specify the assumptions defining the sector considered in the main text. Wedge products between differential forms are understood throughout.

\subsection{Action and normalization}
\label{app:EChS-action}

The five-dimensional Einstein--Chern--Simons action obtained through the S-expansion procedure is \cite{salg1}
\begin{equation}
	S_{\mathrm{EChS}}[e,\omega,h,k]
	=
	\int_{M}
	\epsilon_{abcde}
	\left[
	\alpha_{1}l^{2}R^{ab}R^{cd}e^{e}
	+
	\alpha_{3}
	\left(
	\frac{2}{3}R^{ab}e^{c}e^{d}e^{e}
	+
	2l^{2}k^{ab}R^{cd}T^{e}
	+
	l^{2}R^{ab}R^{cd}h^{e}
	\right)
	\right].
	\label{eq:EChS-action-original-app}
\end{equation}
Here, $e^{a}$ is the vielbein, $\omega^{ab}$ is the spin connection, and $h^{a}$ and $k^{ab}$ are additional bosonic one-form fields. The curvature and torsion two-forms are defined as $R^{ab}=\mathrm{d}\omega^{ab}+\omega^{a}{}_{c}\omega^{cb}$ and $T^{a}=\mathrm{d}e^{a}+\omega^{a}{}_{b}e^{b}$, respectively.

The term proportional to $\alpha_{3}$ in Eq.~\eqref{eq:EChS-action-original-app} contains the five-dimensional
Einstein--Hilbert Lagrangian. Requiring the recovery of Einstein gravity in the limit $l\rightarrow0$ fixes its normalization as
\begin{equation}
	\alpha_{3}
	=
	\frac{1}{8\kappa},
	\label{eq:alpha3-app}
\end{equation}
where $\kappa$ is the five-dimensional gravitational coupling constant. Introducing the dimensionless parameter $\alpha:=\alpha_{1}/\alpha_{3}$, the action becomes
\begin{equation}
	S_{\mathrm{EChS}}[e,\omega,h,k]
	=
	\frac{1}{8\kappa}
	\int_{M}
	\epsilon_{abcde}
	\left(
	\alpha l^{2}R^{ab}R^{cd}e^{e}
	+
	\frac{2}{3}R^{ab}e^{c}e^{d}e^{e}
	+
	2l^{2}k^{ab}R^{cd}T^{e}
	+
	l^{2}R^{ab}R^{cd}h^{e}
	\right).
	\label{eq:EChS-action-app}
\end{equation}
The terms beyond the Einstein--Hilbert contribution are proportional to $l^{2}$. Therefore, the limit $l\rightarrow0$, with $\kappa$ fixed, reduces the theory to five-dimensional Einstein--Hilbert gravity.

\subsection{Field equations in the presence of matter}
\label{app:EChS-matter-equations}

We consider the total action
\begin{equation}
	S
	=
	S_{\mathrm{EChS}}
	+
	S_{\mathrm{M}}
	\label{eq:total-action-app}
\end{equation}
where the matter action may depend on all independent one-form fields and other matter fields. The independent variations of the total action yield
\begin{subequations}
	\label{eq:full-EChS-field-equations-app}
	\begin{align}
		\frac{1}{8}
		\epsilon_{abcde}
		\left(
		\alpha l^{2}R^{bc}R^{de}
		+
		2R^{bc}e^{d}e^{e}
		+
		2l^{2}\mathrm{D}_{\omega}k^{bc}R^{de}
		\right)
		&=
		-\kappa
		\frac{\delta S_{\mathrm{M}}}{\delta e^{a}},
		\label{eq:EChS-e-equation-app}
		\\
		\frac{1}{4}
		\epsilon_{abcde}
		\left(
		\alpha l^{2}R^{cd}T^{e}
		+
		e^{c}e^{d}T^{e}
		+
		l^{2}\mathrm{D}_{\omega}k^{cd}T^{e}
		+
		l^{2}R^{cd}\mathrm{D}_{\omega}h^{e}
		+
		l^{2}R^{cd}k^{e}{}_{f}e^{f}
		\right)
		&=
		-\kappa
		\frac{\delta S_{\mathrm{M}}}{\delta\omega^{ab}},
		\label{eq:EChS-omega-equation-app}
		\\
		\frac{1}{8}
		l^{2}
		\epsilon_{abcde}
		R^{bc}R^{de}
		&=
		-\kappa
		\frac{\delta S_{\mathrm{M}}}{\delta h^{a}},
		\label{eq:EChS-h-equation-app}
		\\
		\frac{1}{4}
		l^{2}
		\epsilon_{abcde}
		R^{cd}T^{e}
		&=
		-\kappa
		\frac{\delta S_{\mathrm{M}}}{\delta k^{ab}}.
		\label{eq:EChS-k-equation-app}
	\end{align}
\end{subequations}

The matter energy-momentum tensor and the spin-density tensor are defined in an orthonormal frame using the Hodge dual operator\footnote{See \autoref{HodgeDual}.} $\star$, by
\begin{subequations}
	\begin{align}
		\frac{\delta S_{\mathrm{M}}}{\delta e^{a}}
		&=
		\hat T_{ba}\star e^{b},
		\label{eq:matter-energy-momentum-app}\\
		\frac{\delta S_{\mathrm{M}}}{\delta\omega^{ab}}
		&=
		\mathcal{S}_{cab}\star e^{c}.
		\label{eq:spin-current-app}
	\end{align}
\end{subequations}
Similarly, we define the tensor associated with the matter source coupled to $h^{a}$ as
\begin{equation}
	\frac{\delta S_{\mathrm{M}}}{\delta h^{a}}
	=
	-\frac{1}{\alpha}
	\hat T_{ba}^{(h)}\star e^{b}.
	\label{eq:h-energy-momentum-app}
\end{equation}
The normalization factor $-1/\alpha$ is introduced so that the curvature-squared contribution in the Einstein-like field equations \eqref{eq:EChS-e-equation-app} can be written directly in terms of $\hat T_{ab}^{(h)}$.

\subsection{Torsionless and spinless sector}
\label{app:EChS-reduced-sector}

For the wormhole configurations studied in the main text, we impose the torsionless condition ($T^{a}=0$) and $k^{ab}=0$. The latter condition defines a consistent sector provided that the matter source coupled to $k^{ab}$ vanishes. Indeed, after imposing $T^a=0$, the $k^{ab}$ equation requires
\begin{equation}
	\frac{\delta S_{\mathrm{M}}}{\delta k^{ab}}
	=
	0.
	\label{eq:k-source-zero-app}
\end{equation}

We also assume that the physical matter has no intrinsic spin, $\mathcal{S}_{cab}=0$ or equivalently
\begin{equation}
	\frac{\delta S_{\mathrm{M}}}{\delta\omega^{ab}}
	=
	0.
	\label{eq:omega-source-zero-app}
\end{equation}

Under these assumptions, the field equations reduce to
\begin{subequations}
	\label{eq:reduced-EChS-field-equations-app}
	\begin{align}
		\frac{1}{8}
		\epsilon_{abcde}
		\left(
		2R^{bc}e^{d}e^{e}
		+
		\alpha l^{2}R^{bc}R^{de}
		\right)
		&=
		-\kappa \hat T_{ba}\star e^{b},
		\label{eq:reduced-e-equation-app}
		\\
		\frac{1}{4}
		l^{2}
		\epsilon_{abcde}
		R^{cd}\mathrm{D}_{\omega}h^{e}
		&=
		0,
		\label{eq:reduced-omega-equation-app}
		\\
		\frac{1}{8}
		l^{2}
		\epsilon_{abcde}
		R^{bc}R^{de}
		&=
		\frac{\kappa}{\alpha}
		\hat T_{ba}^{(h)}\star e^{b},
		\label{eq:reduced-h-equation-app}
		\\
		\frac{\delta S_{\mathrm{M}}}{\delta k^{ab}}
		&=
		0,
		\label{eq:reduced-k-equation-app}
		\\
		T^{a}
		&=
		0.
		\label{eq:reduced-torsion-equation-app}
	\end{align}
\end{subequations}

Equation~\eqref{eq:reduced-e-equation-app} is the equation used to obtain the wormhole matter variables in the main text. It contains the five-dimensional Einstein contribution and the quadratic-curvature EChS correction. Equation~\eqref{eq:reduced-omega-equation-app} constrains the field $h^{a}$, while Eq.~\eqref{eq:reduced-h-equation-app} relates the curvature-squared term to the source coupled to $h^{a}$.

\section{Einstein-like formulation of Einstein--Chern--Simons gravity}
\label{app:EChS-Einstein-like}

In this appendix, we show how the Einstein--Chern--Simons field equations can be rewritten in an Einstein-like form and identify the contribution associated with the additional one-form field $h^{a}$.

From Eq.~\eqref{eq:reduced-h-equation-app}, we have
\begin{equation}
	\frac{1}{8}
	l^{2}
	\epsilon_{abcde}
	R^{bc}R^{de}
	=
	\frac{\kappa}{\alpha}
	\hat T_{ba}^{(h)}
	\star e^{b}.
	\label{eq:h-source-field-equation-app}
\end{equation}
Multiplying this equation by $\alpha$ gives
\begin{equation}
	\frac{1}{8}
	\alpha l^{2}
	\epsilon_{abcde}
	R^{bc}R^{de}
	=
	\kappa
	\hat T_{ba}^{(h)}
	\star e^{b}.
	\label{eq:curvature-h-source-app}
\end{equation}

Substituting Eq.~\eqref{eq:curvature-h-source-app} into the vielbein field equation \eqref{eq:reduced-e-equation-app} yields 
\begin{equation}
	\frac{1}{4}
	\epsilon_{abcde}
	R^{bc}e^{d}e^{e}
	=
	-\kappa
	\left(
	\hat T_{ba}
	+
	\hat T_{ba}^{(h)}
	\right)
	\star e^{b}.
	\label{eq:Einstein-like-equation-app}
\end{equation}

The left-hand side of Eq.~\eqref{eq:Einstein-like-equation-app} is the Einstein tensor expressed in differential-form notation. The EChS field equation can therefore be rewritten as an Einstein-like equation sourced by the effective tensor
\begin{equation}
	\hat T_{ab}^{\mathrm{eff}}
	=
	\hat T_{ab}
	+
	\hat T_{ab}^{(h)}.
	\label{eq:effective-energy-momentum-app}
\end{equation}

This decomposition must be interpreted carefully. The tensor $\hat T_{ab}$ is the ordinary energy-momentum tensor obtained by varying the matter action with respect to the vielbein. By contrast, $\hat T^{(h)}_{ab}$ represents the independent current coupled to the additional one-form $h^a$. Although both quantities arise from the matter action in the present formulation, they correspond to variations with respect to different gauge fields and therefore play distinct roles in the field equations. The standard pointwise energy conditions are imposed on $\hat T_{ab}$, since these conditions are defined for the stress-energy tensor associated with the vielbein. The current $\hat T^{(h)}_{ab}$ is nevertheless part of the complete EChS system and, in the Einstein-like formulation, accounts for the contribution associated with the quadratic-curvature sector. Consequently, the energy conditions examined in the main text are imposed directly on $\hat T_{ab}$ rather than on the effective combination $\hat T^{\rm eff}_{ab}$.

\subsection{Effective source associated with the $h^{a}$ field}
\label{app:h-effective-source}

For the five-dimensional Morris--Thorne geometry, we write the tensor associated with the $h^{a}$ sector in an orthonormal frame as
\begin{equation}
	\hat T^{(h)a}{}_{b}
	=
	\operatorname{diag}
	\left(
	-\rho^{(h)},
	p_{r}^{(h)},
	p_{l}^{(h)},
	p_{l}^{(h)},
	p_{l}^{(h)}
	\right).
	\label{eq:h-source-tensor-app}
\end{equation}
Here, $\rho^{(h)}(r)$, $p_{r}^{(h)}(r)$, and $p_{l}^{(h)}(r)$ denote the energy density, radial pressure, and lateral pressure associated with the $h^{a}$ source, respectively.

Substituting the Morris--Thorne metric \eqref{eq:Morris-Thorne-metric} into Eq.~\eqref{eq:h-source-field-equation-app} gives
\begin{subequations}
	\label{eq:h-source-components-app}
	\begin{align}
		\kappa\rho^{(h)}(r)
		&=
		-\frac{3\alpha l^{2}b(r)
			\big(rb'(r)-b(r)\big)}{2r^{6}},
		\label{eq:h-energy-density-app}
		\\
		\kappa p_{r}^{(h)}(r)
		&=
		\frac{3\alpha l^{2}\Phi'(r)b(r)
			\big(b(r)-r\big)}{r^{5}},
		\label{eq:h-radial-pressure-app}
		\\
		\kappa p_{l}^{(h)}(r)
		&=
		-\frac{\alpha l^{2}}{r^{5}}
		\left\{
		b(r)r
		\big(r-b(r)\big)
		\left(
		\Phi''(r)
		+
		\Phi'(r)^{2}
		\right)
		+
		\Phi'(r)
		\left(r-\frac{3}{2}b(r)\right)
		\big(rb'(r)-b(r)\big)
		\right\}.
		\label{eq:h-lateral-pressure-app}
	\end{align}
\end{subequations}

These expressions coincide, with opposite sign, with the terms proportional to $\alpha l^{2}$ in the equations for the physical matter variables. In particular, combining the physical and $h^{a}$ sectors gives
\begin{equation}
	\kappa
	\left(\rho(r)+\rho^{(h)}(r)\right)
	=
	\frac{3
		\big(rb'(r)+b(r)\big)}{2r^{3}},
	\label{eq:effective-energy-density-app}
\end{equation}
and
\begin{equation}
	\kappa
	\left(p_{r}(r)+p_{r}^{(h)}(r)\right)
	=
	-\frac{3
		\big(
		r\Phi'(r)b(r)
		-
		r^{2}\Phi'(r)
		+
		b(r)
		\big)}{r^{3}}.
	\label{eq:effective-radial-pressure-app}
\end{equation}
For the lateral pressure, one obtains
\begin{equation}
	\kappa
	\left(p_{l}(r)+p_{l}^{(h)}(r)\right)
	=
	\frac{r-b(r)}{r}
	\left(\Phi''(r)
	+
	\Phi'(r)^{2}
	\right)
	+
	\frac{
		\big(
		4r-3b(r)-rb'(r)
		\big)
		\Phi'(r)
	}{2r^{2}}
	-
	\frac{b'(r)}{r^{2}}.
	\label{eq:effective-lateral-pressure-app}
\end{equation}

Equations~\eqref{eq:effective-energy-density-app}--\eqref{eq:effective-lateral-pressure-app} reproduce the five-dimensional Einstein equations for the Morris--Thorne geometry with the effective source
\begin{equation}
	\hat T_{ab}^{\mathrm{eff}}
	=
	\hat T_{ab}
	+
	\hat T_{ab}^{(h)}.
	\label{eq:effective-source-wormhole-app}
\end{equation}

The Einstein-like representation therefore absorbs the quadratic-curvature correction into the additional tensor $\hat T_{ab}^{(h)}$. Nevertheless, this is only a reformulation of the field equations. The physical matter tensor remains $\hat T_{ab}$, whereas $\hat T^{(h)}_{ab}$ represents the independent current associated with the field $h^a$ and, in the Einstein-like description, accounts for the higher-curvature contribution.

This distinction is central to the analysis in the main text. The aim is to determine whether the physical matter tensor $\hat T_{ab}$ satisfies the standard energy conditions, rather than to transfer their violation to the effective contribution $\hat T_{ab}^{(h)}$.


\end{document}